\documentclass[twocolumn]{aastex631}

\expandafter\def\csname editcolor1\endcsname{magenta}
\expandafter\def\csname editcolor2\endcsname{blue}  
\expandafter\def\csname editcolor3\endcsname{violet} 

\def\chris#1{{\colorcount=2\color{\csname editcolor\the\colorcount\endcsname}
{\ifturnofftwo\else\bf (Chris: #1)\ \fi}}}

\def\jing#1{{\colorcount=1\color{\csname editcolor\the\colorcount\endcsname}
{\ifturnoffone\else\bf (Jing: #1)\ \fi}}}

\def\ben#1{{\colorcount=3\color{\csname editcolor\the\colorcount\endcsname}
{\ifturnoffone\else\bf (Ben: #1)\ \fi}}}

\newcommand{\eg}{e.g.,\ }

\newcommand{\kms}{km~s$^{-1}$}

\newcommand{\SiII}{Si~{\sc ii}}

\newcommand{\CaII}{Ca~{\sc ii}}

\newcommand{\Nifs}{$^{56}$Ni}

\newcommand{\DmB}{\Delta{\rm m_{15}(B)}}

\usepackage{newtxtext,newtxmath}
\usepackage[T1]{fontenc}
\usepackage{graphicx}	
\usepackage{amsmath}	
\usepackage{float}
\usepackage{bookmark} 
\bookmarksetup{numbered, open,}
\usepackage{enumitem}
\setlist[enumerate]{itemsep=0mm}
\usepackage{hyperref}
\usepackage{subfigure}
\usepackage{xspace}
\usepackage{natbib}
\usepackage{xcolor}  

\graphicspath{{./}{figures/}}

\received{11th April 2022 }
\revised{12th May 2022}
\accepted{22nd May 2022}

\submitjournal{ApJL}

\shorttitle{Early excess emission in SN 2021aefx}
\shortauthors{C. Ashall, J.~Lu, et al.}

\begin{document}

\title{A Speed Bump: SN~2021aefx Shows that Doppler Shift Alone can Explain Early-Excess Blue Flux in Some Type Ia Supernovae}

\correspondingauthor{Chris~Ashall}
\email{chris.ashall24@gmail.com}
\author[0000-0002-5221-7557]{C.~Ashall}
\affil{Institute for Astronomy, University of Hawaii, 2680 Woodlawn Drive, Honolulu, HI 96822,USA}

\author[0000-0002-3900-1452]{J.~Lu}
\affil{Department of Physics, Florida State University, 77 Chieftan Way, Tallahassee, FL 32306, USA}

\author[0000-0003-4631-1149]{B.~J.~Shappee}
\affil{Institute for Astronomy, University of Hawaii, 2680 Woodlawn Drive, Honolulu, HI 96822,USA}

\author[0000-0003-4625-6629]{C.~R.~Burns}
\affiliation{The Observatories of the Carnegie Institution for Science, 813 Santa Barbara Street, Pasadena, CA 91101, USA}

\author[0000-0003-1039-2928]{E.~Y.~Hsiao}
\affil{Department of Physics, Florida State University, 77 Chieftan Way, Tallahassee, FL 32306, USA}

\author[0000-0001-8367-7591]{S.~Kumar}
\affil{Department of Physics, Florida State University, 77 Chieftan Way, Tallahassee, FL 32306, USA}

\author[0000-0003-2535-3091]{N.~Morrell}
\affil{Carnegie Observatories, Las Campanas Observatory, Casilla 601, La Serena, Chile}

\author[0000-0003-2734-0796]{M.~M.~Phillips}
\affil{Carnegie Observatories, Las Campanas Observatory, Casilla 601, La Serena, Chile}

\author[0000-0002-9301-5302]{M.~Shahbandeh}
\affil{Department of Physics, Florida State University, 77 Chieftan Way, Tallahassee, FL 32306, USA}

\author[0000-0001-5393-1608]{E.~Baron}
\affiliation{Homer L. Dodge Department of Physics and Astronomy, University of Oklahoma, 440 W. Brooks, Rm 100, Norman, OK 73019-2061, USA}
\affiliation{Hamburger Sternwarte, Gojenbergsweg 112, D-21029 Hamburg, Germany}
\affiliation{Department of Physics, The George Washington University, Corcoran Hall, 725 21st Street NW, Washington, DC 20052, USA}

\author[0000-0003-4432-5037]{K.~Boutsia}
\affiliation{Carnegie Observatories, Las Campanas Observatory, Casilla 601, La Serena, Chile}

\author[0000-0001-6272-5507]{P.~J.~Brown}
\affiliation{George P. and Cynthia Woods Mitchell Institute for Fundamental
 Physics and Astronomy, Department of Physics and Astronomy, Texas
 A\&M University, College Station, TX 77843, USA}

\author[0000-0002-7566-6080]{J.~M.~DerKacy}
\affiliation{Homer L. Dodge Department of Physics and Astronomy, University of Oklahoma, 440 W. Brooks, Rm 100, Norman, OK 73019-2061, USA}

\author[0000-0002-1296-6887]{L.~Galbany}
\affiliation{Institute of Space Sciences (ICE, CSIC), Campus UAB, Carrer de Can Magrans, s/n, E-08193 Barcelona, Spain}
\affiliation{Institut d’Estudis Espacials de Catalunya (IEEC), E-08034 Barcelona, Spain}

\author[0000-0002-4338-6586]{P.~Hoeflich}
\affiliation{Department of Physics, Florida State University, 77 Chieftan Way, Tallahassee, FL 32306, USA}

\author[0000-0002-6650-694X]{K.~Krisciunas}
\affiliation{George P. and Cynthia Woods Mitchell Institute for Fundamental
 Physics and Astronomy, Department of Physics and Astronomy, Texas
 A\&M University, College Station, TX 77843, USA}

\author{P. Mazzali}
\affil{Astrophysics Research Institute, Liverpool John Moores University, IC2, Liverpool Science Park, 146  Brownlow Hill, Liverpool L3 5RF, UK}

\author[0000-0001-6806-0673]{A.~L.~Piro}
\affil{The Observatories of the Carnegie Institution for Science, 813 Santa Barbara Street, Pasadena, CA 91101, USA}

\author[0000-0002-5571-1833]{M.~D.~Stritzinger}
\affiliation{Department of Physics and Astronomy, Aarhus University, Ny Munkegade 120, DK-8000 Aarhus C, Denmark}

\author[0000-0002-8102-181X]{N.~B.~Suntzeff}
\affiliation{George P. and Cynthia Woods Mitchell Institute for Fundamental
 Physics and Astronomy, Department of Physics and Astronomy, Texas
 A\&M University, College Station, TX 77843, USA}

\begin{abstract}
We present early-time photometric and spectroscopic observations of the Type Ia Supernova (SN~Ia) 2021aefx. The early time $u$-band light curve shows an excess flux when compared to normal SNe~Ia. We suggest that the early-excess blue flux may be due to a rapid change in spectral velocity in the first few days post explosion,  produced by the emission of the \CaII\ H\&K feature passing from the $u$ to the $B$ bands on the time scale of a few days. This effect could be dominant for all SNe~Ia which have broad absorption features and early-time velocities over 25,000\,\kms. It is likely to be one of the main causes of early-excess u-band flux in SNe Ia which have early-time high-velocities. This effect may also be dominant in the UV filters, as well as in places where the SN spectral energy distribution is quickly rising to longer wavelengths. The rapid change in velocity can only produce a monotonic change (in flux-space) in the $u$-band. For objects which explode at lower velocities, and have a more structured shape in the early-excess emission, there must also be an additional parameter producing the early-time diversity. More early time observations, in particular early spectra, are required to determine how prominent this effect is within SNe~Ia.
\end{abstract}


\section{Introduction}
Type Ia Supernovae (SNe~Ia) are thought to come from the thermonuclear explosion of at least one Carbon-Oxygen White Dwarf (WD) in a binary system  \citep{Whelan73,Iben84}. Their peak luminosities are commonly used to measure luminosity distances to define the Hubble flow \citep[\eg][]{Phillips93,Riess98,Perlmutter99}.
Yet, to date, determining the exact progenitor scenario (e.g. single degenerate vs double degenerate) or explosion mechanism (Chandrasekhar mass vs sub-Chandrasekhar mass) has eluded the community \citep[\eg][]{Maoz14,Blondin17,Hoeflich17,Ashall18,Galbany19,Jha19}.

In recent years high-cadence transient surveys such as the All-Sky Automated Survey for Supernovae (ASAS-SN; \citealt{Shappee14,Kochanek17}), the Asteroidal Terrestrial-impact Last Alert System (ATLAS; \citealt{Tonry18}), the Distance less than 40 (DLT40)\footnote{http://dark.physics.ucdavis.edu/dlt40/DLT40}, and the Zwicky Transient Facility (ZTF; \citealt{Bellm19}), have dramatically increased the number of SNe~Ia caught within hours to days of explosion. The radiative transfer through the outer layers which can only be seen up to a few days after explosion allows us to model the explosion and measure the chemical properties of the progenitor system.

There have only been a handful of SNe~Ia that have multi-band photometry and were caught within hours of first light. A subset of these objects are thought to have early-excess blue emission, two-component rising light curves, or blue bumps in their light curves.  
Some of these SNe are: SN~2012cg \citep{Marion16,Shappee18}, SN~2017cbv \citep{Hosseinzadeh17}, ASASSN-18bt/2018oh \citep{Shappee19,Dimitriadis19},  SN~2019yvq \citep{Miller20,Tucker21},  and SN~2020hvf \citep{Jiang21}.  For a summary of SNe~Ia with early-excess light-curve emission, see \citet{Jiang18}.  Another subset of SNe~Ia have smooth rising light curves such as SN~2011fe \citep{Nugent11}, ASASSN-14lp \citep{Shappee16}, and SN~2015F \citep{Cartier17}. Furthermore, the early-time color curves of SNe~Ia have  shown diversity \citep{Stritzinger18,Bulla20}.

The cause of the early-time light curve diversity is unknown, but various effects have been proposed. These include: 1) the collision of the ejecta with a companion star in a single degenerate system \citep{Marietta00,Kasen10}; 2) the presence of \Nifs\ in the outer layer, which can be mixed out from the center of the explosion or produced {\em in situ}  \citep{Piro16,Ni22}; or 3) the interaction of the ejecta with circumstellar material  \citep{Gerardy07,Dragulin16,Piro16, Jiang21}.

However, to date, no work has been carried out on determining if rapidly changing spectral features can produce the observed diversity in the rising light curves. In this work, we  quantify the effect that a rapidly changing spectral feature has on the fixed-filter  early light curves. We use observations of SN~2021aefx as a test case. 
In section \ref{sect:data} the data are presented, in section \ref{sect:Treatment} we discuss how the data are calibrated, in section \ref{sect:results} the results are shown, and the discussion is presented in section \ref{sect:Discussion}. 

\section{Data}\label{sect:data}

\subsection{Discovery}
SN 2021aefx was discovered by the DLT40 transient survey on 2021-11-11T12:30:43.78 (MJD 59529.52) at 17.24~mag, with a last non-detection on 2021-11-06T07:52:17 (MJD 59524.33) at 19.35~mag.  Both the discovery and last non-detection were in a clear filter \citep{DLT40}. SN 2021aefx was classified as a high-velocity SN~Ia  using a spectrum obtained $\sim$8 hours (2021-11-11T20:55:27; MJD 59529.87) after discovery \citep{classification}. ASAS-SN (\citealt{Shappee14,Kochanek17}) places tighter constraints on the last non-detection with a limiting 3-sigma observation of 17.70~mag in the $g$-band on 2021-11-09T08:37:39 (MJD 59527.36). The host of SN 2021aefx (NGC~1566) has a heliocentric redshift of 0.0050 \citep{Allison14}, which corresponds to a distance modulus of  31.43$\pm$0.15\,mag when corrected for the local peculiar motion of the Virgo, Great Attractor, and Shapley clusters \citep{Mould00}. Alternatively, using the Tip of the Red Giant Branch (TRGB) produces a distance modulus of 31.27$\pm$0.49\,mag \citep{Sabbi18}.

\begin{figure}
    \centering
    \includegraphics[width=0.98\columnwidth]{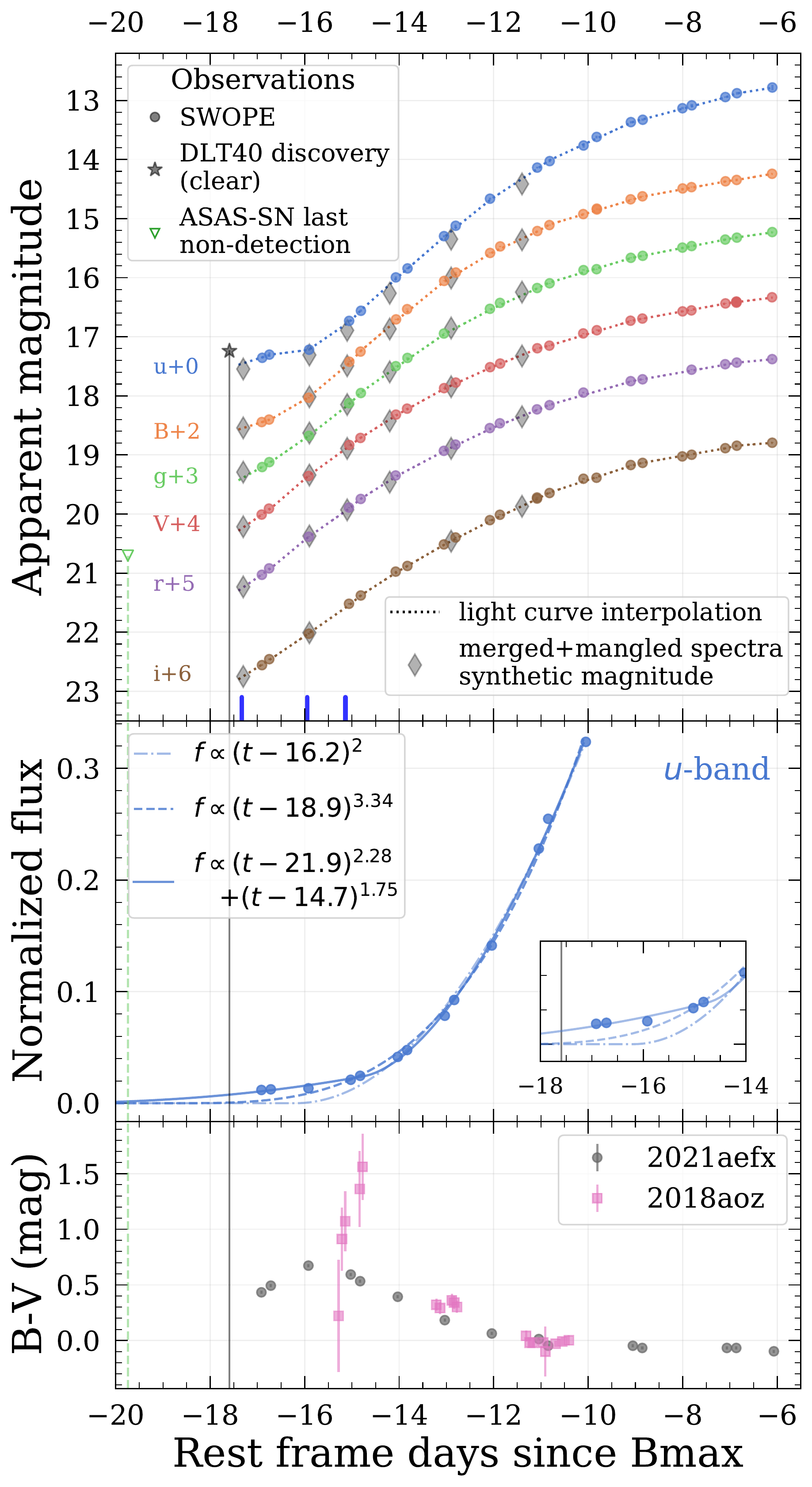} 
    \caption{(Top:) Multi-band photometric observations of SN~2021aefx. The ASAS-SN last non-detection and DLT40 discovery are indicated with open triangle and filled star, respectively. The dotted lines show the interpolation of the early time light curves that is used for color correcting the spectra. The diamonds represent the synthetic magnitudes of integrating the mangled spectra described in Section~\ref{sect:Treatment}. The vertical blue lines on the bottom mark the epochs of the observed spectra of SN~2021aefx. (Middle:) Power-law fits to the rising $u$-band light. The dash-dotted line is the f$\propto$t$^{2}$ fit, the dashed line allows for the power-law index to vary, and the solid line is the two component fit. (Bottom:) The early-time $B-V$ color curves of SN~2021aefx and SN~2018aoz. Both SNe have an initial period where they turn red before becoming blue. The error bars are plotted but are smaller than the points.
    The plotted light- and color-curves are corrected for Milky-Way extinction.}
    \label{fig:LC_21aefx}
\end{figure}

\subsection{Photometry}
 Multi-band ($uBVgri$) follow-up observations were initiated by the Precision Observations for Infant Supernovae Explosions (POISE, \citealt{POISE}) a day after discovery. The photometry was obtained on the 1m Swope telescope at Las Campanas Observatory. Observations were acquired twice per night in order to look for small scale fluctuations in the light curve. The photometry was reduced and analyzed using the same methods outlined by \citet{Krisciunas17} and \citet{Phillips19}. The uncertainties shown in Fig. 1 are derived from the usual propagation of errors assuming Poisson statistics, as well as the uncertainty in the nightly zero-points derived from observations of standard stars (see \citealt{Krisciunas17}). A further source of potential error which is not included is the contamination of host galaxy light. While visual inspection indicates this is likely very small, final photometry and errors will require observations of the host galaxy after the SN has faded. A log of the photometric observations can be found in Table \ref{table:log_lc}. The early time light curve is presented in the top panel of Fig. \ref{fig:LC_21aefx}. The middle panel in Fig. \ref{fig:LC_21aefx} demonstrates that the rise is not well characterized by a simple constant-temperature fireball expansion f $\propto$ t$^2$, which gives an invalid time of explosion that is after the discovery. However, there is no good reason why the f $\propto$ t$^2$ should provide an exact match to the data \citep{Nugent11}. 
A power-law that allows the index to vary (f $\propto$ t$^{3.34}$) fits the $u$-band rise better, but still underestimates the flux of the first three data points.
In fact, the $u$-band light curve may be best described as having a two-component rise, which may be identified as early-excess emission in the two days following discovery. For the two component fit and the fireball model the 13 data points shown in the middle panel of Fig. \ref{fig:LC_21aefx} used in the fitting process.
The two-component fit has six free parameters and a reduced $\chi^2$ of 28.
Whereas  the fireball model  had two free parameters and a  reduced $\chi^2$ of 664, and the one-component power-law fit has three free parameters and reduced $\chi^2$ is 276.
We note that the two component rise in SN~2021aefx is different from the one seen in the Kepler filter for ASASSN-18bt \citep{Shappee19}. This is because the Kepler filter covers a much larger wavelength range ($\sim$4000-9000\AA). Therefore, in SN~2021aefx the effect likely comes from a different physical process. 

A full detailed analysis of the later epochs of SN~2021aefx will be presented in a future paper. However, here we provide some important information about the SN for context. SN~2021aefx peaked at an apparent  $B$-band magnitude of 11.99$\pm$0.01\,mag, and has a decline rate of $\DmB$=1.01 $\pm$ 0.06 mag.\footnote{$\DmB$ is the change in $B$-band magnitude between maximum light and 15~days past maximum.} This corresponds to an absolute $B$-band magnitude of --19.28$\pm$0.49\,mag, using the TRGB distance of  $\mu$=31.27\,mag, and a foreground Galactic extinction of $E(B-V)_{MW}$=0.01\,mag \citep{Schlafly11}.
Throughout this work the data have been correct for foreground Galactic extinction. This places it in the normal part of the luminosity width relation, and in a similar location to SN~2009ig, which has $M_B$= --19.46$\pm$0.50\,mag and $\DmB$ = 0.90$\pm$0.07\,mag \citep{Marion13}.

Furthermore, the $B-V$ color curve of SN\,2021aefx has the unusual behavior where it starts red, and then turns blue within the first two days, see Fig \ref{fig:LC_21aefx}. This is unlike any SN in the $B-V$ sample of \citet{Stritzinger18}, but is similar to SN~2018aoz.  However,  the evolution of SN~2018aoz is more extreme compared to  SN~2021aefx. For SN~2018aoz this effect has been interpreted as enhanced line blanketing in the outer layers of the ejecta caused by iron group elements \citep{Ni22}.

\begin{deluxetable*}{cccccccc}[htb!]
\centering
\tablecaption{Swope e2v optical photometry of SN~2021aefx.\tablenotemark{a}\label{table:log_lc}}
\tablehead{
\colhead{MJD} & \colhead{Phase\tablenotemark{b}} & \colhead{$u$} & \colhead{$g$} & \colhead{$B$} & \colhead{$V$} & \colhead{$r$} & \colhead{$i$}}
\startdata
59530.2	&	$-$16.9	&	17.39(0.02)	&	16.48(0.01)	&	16.24(0.01)	&	16.04(0.01)	&	16.05(0.01)	&	16.57(0.01)	\\
59530.4	&	$-$16.7	&	17.34(0.04)	&	16.43(0.01)	&	16.15(0.01)	&	15.93(0.01)	&	15.94(0.01)	&	16.47(0.01)	\\
59531.2	&	$-$15.9	&	17.26(0.02)	&	16.06(0.01)	&	15.71(0.01)	&	15.38(0.01)	&	15.41(0.01)	&	16.04(0.01)	\\
59532.1	&	$-$15.0	&	16.77(0.02)	&	15.46(0.01)	&	15.15(0.01)	&	14.86(0.01)	&	14.91(0.01)	&	15.54(0.01)	\\
59532.3	&	$-$14.8	&	16.60(0.02)	&	15.28(0.01)	&	14.98(0.01)	&	14.74(0.01)	&	14.77(0.01)	&	15.40(0.02)	\\
59533.1	&	$-$14.0	&	16.03(0.02)	&	14.74(0.01)	&	14.53(0.01)	&	14.34(0.01)	&	14.37(0.01)	&	14.99(0.01)	\\
59533.3	&	$-$13.8	&	15.88(0.02)	&	14.56(0.01)	&	14.39(0.01)	&	14.24(0.01)	&	$\cdots$	&	14.90(0.01)	\\
59534.1	&	$-$13.0	&	15.34(0.02)	&	14.09(0.01)	&	13.98(0.01)	&	13.90(0.01)	&	13.95(0.01)	&	14.53(0.01)	\\
59534.3	&	$-$12.8	&	15.16(0.02)	&	13.94(0.01)	&	$\cdots$	&	13.8(0.01)	&	13.85(0.01)	&	14.41(0.01)	\\
59535.1	&	$-$12.0	&	14.70(0.02)	&	13.61(0.01)	&	13.56(0.01)	&	13.54(0.01)	&	13.57(0.01)	&	14.12(0.01)	\\
59535.3	&	$-$11.8	&	$\cdots$	&	13.50(0.01)	&	13.46(0.01)	&	13.48(0.01)	&	13.49(0.01)	&	14.03(0.01)	\\
59536.1	&	$-$11.0	&	14.18(0.02)	&	13.24(0.01)	&	13.20(0.01)	&	13.22(0.01)	&	13.25(0.01)	&	13.75(0.01)	\\
59536.3	&	$-$10.8	&	14.06(0.02)	&	13.14(0.01)	&	13.12(0.01)	&	13.18(0.01)	&	13.18(0.01)	&	13.66(0.01)	\\
59537	&	$-$10.1	&	$\cdots$	&	12.96(0.01)	&	$\cdots$	&	12.97(0.01)	&	12.97(0.01)	&	13.42(0.01)	\\
59537.1	&	$-$10.0	&	13.80(0.04)	&	$\cdots$	&	12.90(0.01)	&	$\cdots$	&	$\cdots$	&	$\cdots$	\\
59537.3	&	$-$9.9	&	13.66(0.02)	&	12.87(0.01)	&	12.89(0.01)	&	12.92(0.01)	&	$\cdots$	&	13.40(0.01)	\\
59538.1	&	$-$9.1	&	13.40(0.02)	&	12.71(0.01)	&	12.69(0.01)	&	12.75(0.01)	&	12.77(0.01)	&	13.19(0.01)	\\
59538.3	&	$-$8.9	&	13.36(0.02)	&	12.66(0.01)	&	12.66(0.01)	&	12.72(0.01)	&	12.74(0.01)	&	13.15(0.01)	\\
59539.2	&	$-$8.0	&	13.17(0.02)	&	12.52(0.01)	&	12.52(0.01)	&	12.60(0.01)	&	$\cdots$	&	13.04(0.01)	\\
59539.3	&	$-$7.9	&	$\cdots$	&	12.50(0.01)	&	$\cdots$	&	12.58(0.01)	&	12.58(0.01)	&	$\cdots$	\\
59539.4	&	$-$7.8	&	13.12(0.02)	&	$\cdots$	&	12.50(0.01)	&	$\cdots$	&	$\cdots$	&	13.01(0.01)	\\
59540.1	&	$-$7.1	&	12.98(0.01)	&	12.40(0.01)	&	12.38(0.01)	&	12.46(0.01)	&	12.49(0.01)	&	12.90(0.01)	\\
59540.3	&	$-$6.9	&	12.92(0.02)	&	12.38(0.01)	&	12.35(0.01)	&	12.44(0.01)	&	12.46(0.01)	&	12.86(0.01)	\\
59541.1	&	$-$6.1	&	12.82(0.05)	&	12.27(0.01)	&	12.26(0.01)	&	12.36(0.01)	&	12.40(0.01)	&	12.81(0.01)	\\
\enddata
\tablenotetext{a}{Magnitudes in the Swope+e2v natural system.}
\tablenotetext{b}{Relative to the $B$-band maximum of 2021aefx at MJD=59547.29.}
\end{deluxetable*}

\subsection{Spectroscopy}
\begin{figure*}
    \centering
    \includegraphics[width=0.98\textwidth]{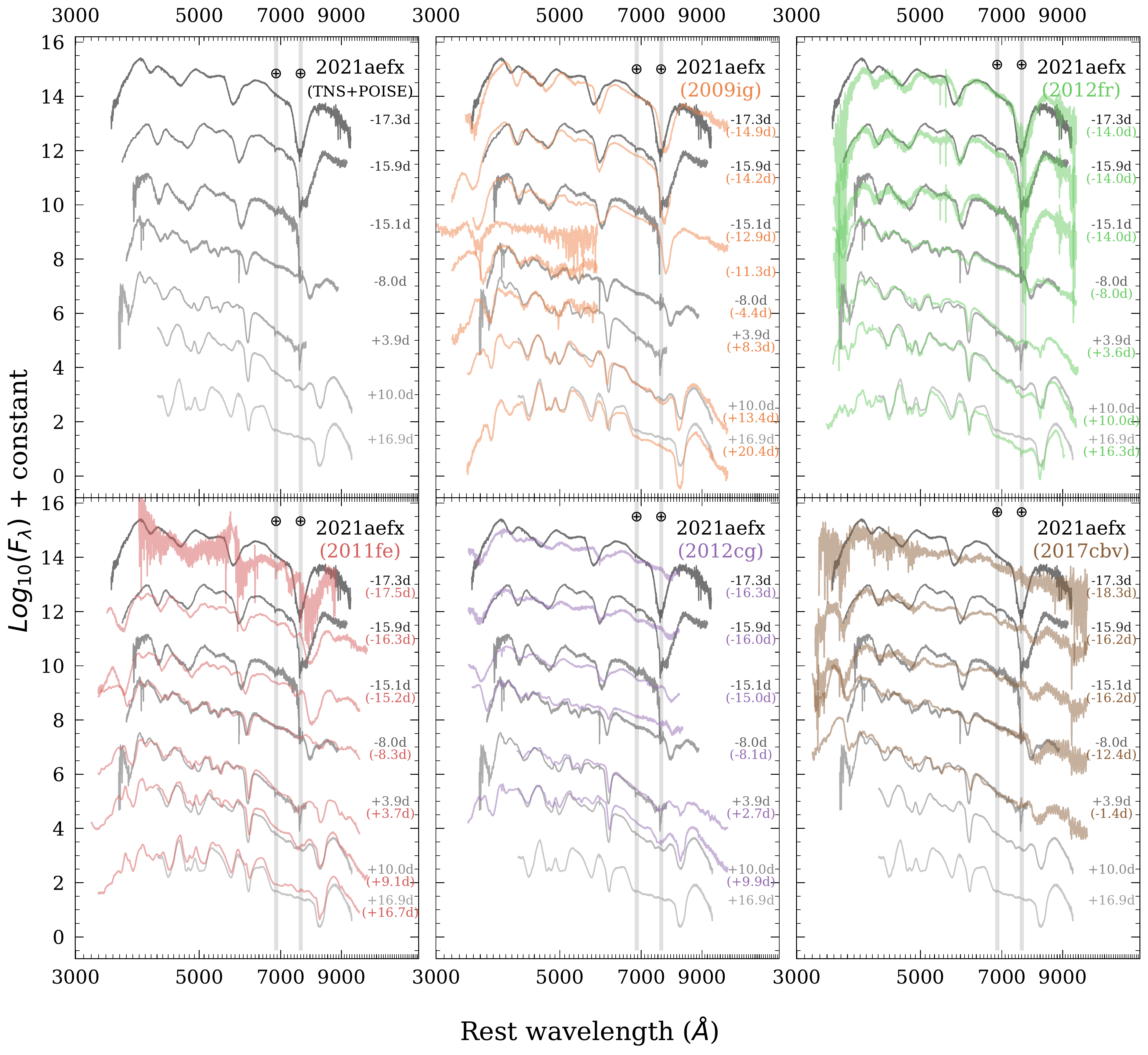} 
    \caption{Spectra comparison of SN~2021aefx and a variety of other SNe~Ia.  SN~2021aefx is most similar to  SNe~2009ig and 2012fr. Interestingly, the comparison also shows that 2021aefx is not the same as SN~2012cg or SN~2017cbv despite them having an early-excess blue flux. 
    Throughout this work, we use the spectra of SN~2009ig to fill in gaps in wavelength and phase. }
    \label{fig:spec_compare_by_sn}
\end{figure*}

We  present seven spectra of SN 2021aefx. Four were taken from the Transient Name Server database,\footnote{https://www.wis-tns.org/} one was acquired using MIKE on the 6.5~m Magellan Clay Telescope, and two were obtained using IMACS on the 6.5~m Magellan Baade Telescope at Las Campanas Observatory.  
The spectra were reduced using
standard \textsc{iraf}\footnote{The Image Reduction and Analysis Facility (iraf) is distributed
by the National Optical Astronomy Observatory, which is operated by the Association of Universities for Research in Astronomy, Inc., under cooperative agreement with the National Science
Foundation.} packages following the method as described by \citet{Hamuy06}. A log of the spectra can be found in Table~\ref{tab:log_spec}.  The spectra are presented in Fig. \ref{fig:spec_compare_by_sn}.

\begin{table}[tbh!]
\centering
\tabletypesize{\small}
\caption{Journal of Spectroscopic Observations.}
\addtolength{\tabcolsep}{-4pt}
\label{tab:log_spec}
\begin{tabular}{c|c|c|c|c}
\hline\hline
UT Date     & MJD   & Phase\tablenotemark{a}  & Telescope/Instrument      & Source \\
 & & [day] & & \\ 
\hline
2021-11-11 & 59529.87 & -17.3 & SALT + RSS         & TNS          \\
2021-11-13 & 59531.27 & -15.9 & NTT + EFOSC2       & ePESSTO+     \\
2021-11-14 & 59532.08 & -15.1 & Alpy600 + 35cm R-C              & TNS          \\
2021-11-21 & 59539.29 & -7.9  & Clay + MIKE        & POISE        \\                 
2021-12-03 & 59551.21 & 3.9   & Alpy600 + 35cm R-C              & TNS          \\
2021-12-09 & 59557.35 & 10.0  & Baade + IMACS      & POISE        \\
2021-12-16 & 59564.27 & 16.9  & Baade + IMACS      & POISE        \\
\hline
\end{tabular}
\tablenotetext{a}{Relative to the $B$-band maximum of 2021aefx at MJD=59547.29.}
\end{table}

\begin{figure*}
    \centering
    \includegraphics[width=0.98\textwidth]{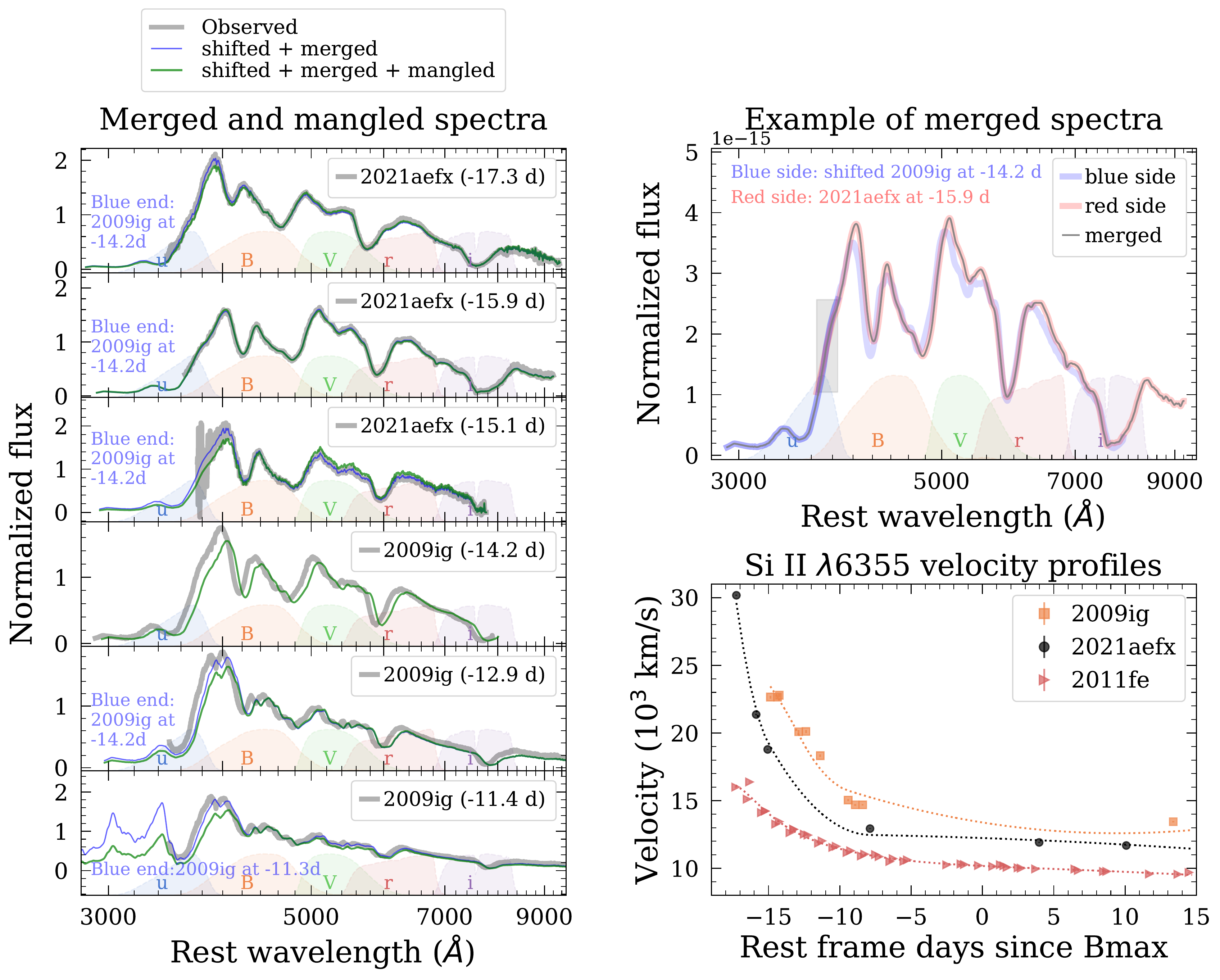} 
    \caption{(Left:) The six color-adjusted spectra chosen for this analysis, including three early spectra of SN~2021aefx and three spectra of SN~2009ig. The spectra are color-matched to the observed photometry of SN~2021aefx. (Top right:) An example of the merging process between SN~2021aefx and SN~2009ig.  Before the merging process, the spectra of SN~2009ig are shifted in velocity to match the velocity of SN~2021aefx at the same phase. The gray shade marks the overlapping region used to match the flux scale.
    (Bottom right:) Velocity profiles of the minimum of the \SiII\ $\lambda$6355 feature in SN~2011fe, SN~2009ig, and SN~2021aefx.}
    \label{fig:merge_spec}
\end{figure*}


The earliest spectrum of SN~2021aefx (at --17.3\,d\footnote{Throughout this work, all phases (in days) are provided relative to rest frame $B$-band maximum.}) was obtained during the early-excess emission phase in the $u$-band light curve.  This early-time spectrum shows high-velocity and broad spectral features. Whereas, by --7.9~d  the spectrum resembles a normal SN Ia. At this phase it is located between the Core normal and Broad Line SNe in  the Branch diagram \citep{Branch09}. At --17.3~d, the velocity of the absorption minimum of the \SiII\ $\lambda$6355 feature reaches $\sim$30,000\,\kms (see bottom right panel of Fig. \ref{fig:merge_spec}). Within 1.4~d, this velocity drops to  $\sim$21,000\,\kms. This deceleration is  larger than other SNe~Ia caught in these early stages, for example SN~2011fe declined from 16,000\,\kms\ to 13,000\,\kms\ over similar epochs. In fact, at early times all of the spectral lines in SN 2021aefx are higher in velocity compared to  SN~2011fe, as is seen in Fig. \ref{fig:spec_compare_by_sn}.

Although it is  apparent that there is diversity in the rising light curves of SNe~Ia,  few spectra have been obtained or analyzed during these early phases. SN 2021aefx is the first public example of a SN with an early excess $u$-band emission which also has high velocity spectra in the early phases. Given that the velocity of the spectral lines rapidly decreases over the course of a few days, it is natural to ask the question: \textit{Could a rapid change in velocity of the ejecta produce the early excess u-band emission?} For example, it could be the case that at early times the spectral features are so shifted to the blue that they affect the broad-band photometry. The blue shift may \textit{add} flux to the $u$-band. 
We investigate this further below. In fact, the sensitivity of broad-band photometry to spectral features near the edges of the photometric bandpasses was noted as far back as \citet{Suntzeff88} when comparing CTIO and SAAO $I$-band photometry. which could differ by 0.3 mag.

\section{Constructing spectrophotometric data} \label{sect:Treatment}
While the spectra cover most of the $u$ band, they do not cover the entire bandpass so careful consideration is warranted. Therefore, assumptions must be made about the nature of the SED in this region. In order to establish which objects have the same ionization state and line ratios,
SN 2021aefx is compared to a range of SNe~Ia which have early-time spectra. This is done to ensure that any extrapolation to the blue is made with a SN which has similar ejecta properties.  At early times, 2021aefx does not resemble SN~2011fe, SN~2012cg, or SN~2017cbv, but is remarkably similar to SN~2012fr \citep{Contreras18} and SN~2009ig \citep{Foley12,Marion13}. Figure \ref{fig:spec_compare_by_sn} shows these comparisons.  The latter objects are SNe with high-velocities and broad features at early times. For example, at $\sim$--15~d the pseudo equivalent width (pEW) of the \SiII\ 6355 feature for SN~2021aefx, SN~2009ig and SN~2012fr is larger than 150\,$\AA$, whereas the pEW of SN~2011fe is $\sim$100\,$\AA$. Conveniently,  some early-time spectra of SN~2009ig extend into the blue and cover the whole wavelength range of the $u$-band.

In order to extend the spectra of SN~2021aefx to cover the whole of the $u$-band region, we \textit{merge} the blue part of the spectra of SN~2009ig onto the spectra of SN~2021aefx. However, before doing this, the spectra of  SN~2009ig are matched to the closest phase of SN~2021aefx, and then shifted in velocity to match the optical features of SN~2021aefx. The assumption made here is that if SNe~Ia are very similar in the $BVgri$ bands, they should also be similar in the $u$-band.
We do note that SNe which are optical twins  might not always have similar mid and far-UV flux. For example, SNe 2011fe and 2011by were spectroscopic twins in the optical but differed in the  mid and far UV \citep{Foley13,Graham15}. However, when the spectra are normalized in the 4000-5500\AA\ region at --10\,d  SNe~2011fe and 2011by only differ in $u$-band flux by 6\%.

An example of the \textit{merged} spectrum is presented in the top right panel of Figure \ref{fig:merge_spec}. 
SN~2009ig does not have $u$-band photometry early enough to determine if it has an early-excess blue flux. However, at the earliest time 96\% of the $u$-band flux in the merged spectrum region comes from the data of SN~2021aefx, and only 4\% comes from the \textit{added/merged} data. To determine if the choice  of SN for the merging process affects the results, we also ran the trial using SN~2011fe as the merged spectrum. Using SN~2011fe only 9\% of the flux in the $u$-band comes from SN~2011fe.  Therefore,  the merging process does not significantly affect our results.

The \textit{merged} spectra are then color matched (also referred to as mangled) with the $uBVgri$ bands  to the observed photometry of SN~2021aefx (see the left panel of Fig. \ref{fig:merge_spec}). 
 The color matching is done using the \textsc{mangle} function in SuperNovae in object oriented Python (SNooPy) program \citep{Burns11}.  
 Cubic splines were fit with control points located at the effective wavelengths of each filter. In addition, two "stabilizing" filters are introduced 100\,\AA\ to the blue of the bluest filter and to the red of the reddest filter. These extra control points are given the same flux as through the bluest and reddest filters, forcing the spline to have zero slope at the ends. Integrating the spectra through the $uBVgri$ filters at each epoch correctly produces the observed light curve, as is seen in  Fig. \ref{fig:LC_21aefx}.

\section{The effects of rapidly changing velocity profiles} \label{sect:results}
\begin{figure*}
    \centering
    \begin{minipage}[c]{.45\linewidth}
        \includegraphics[width=\linewidth]{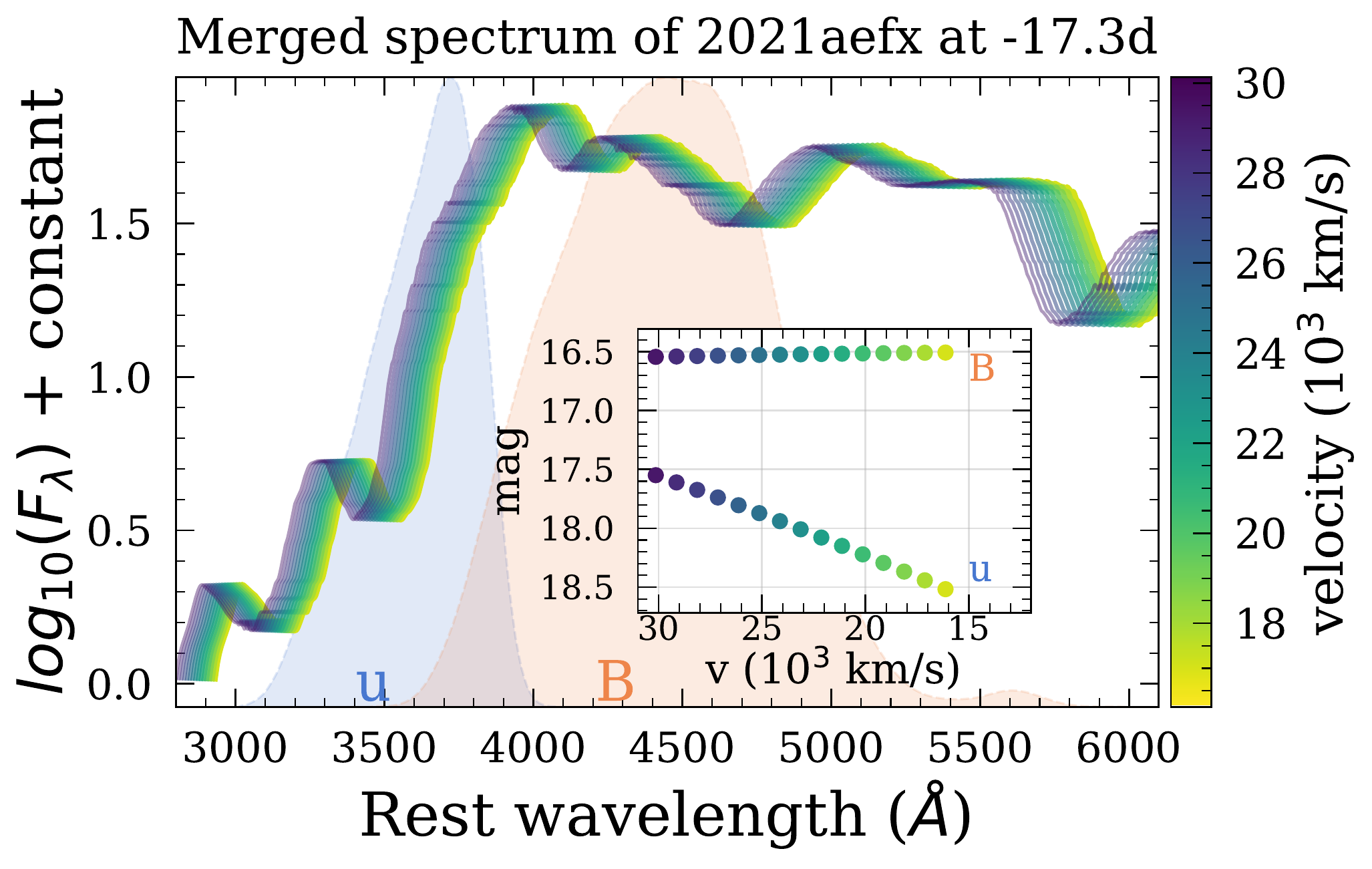} 
        \includegraphics[width=\linewidth]{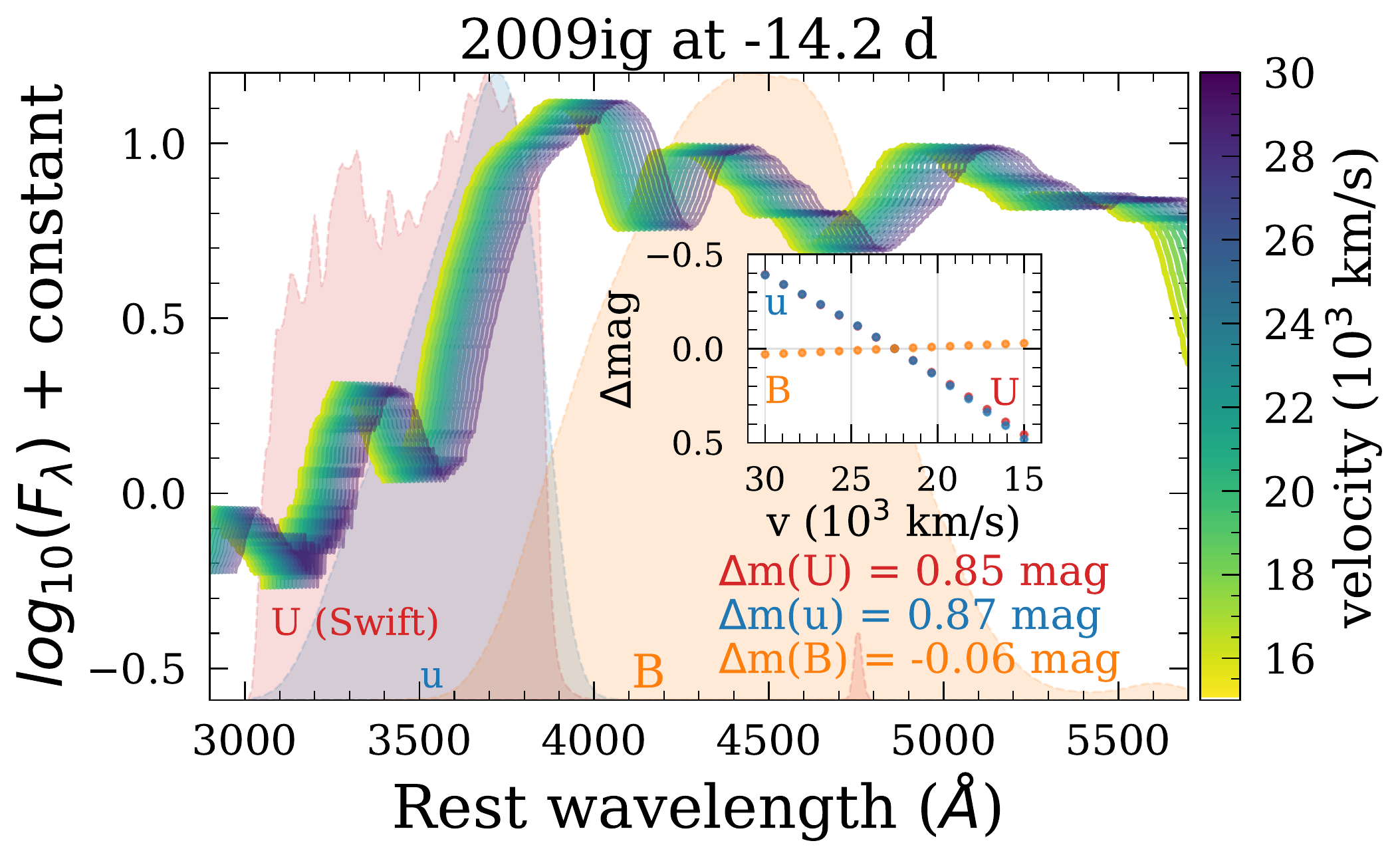} 
        \includegraphics[width=\linewidth]{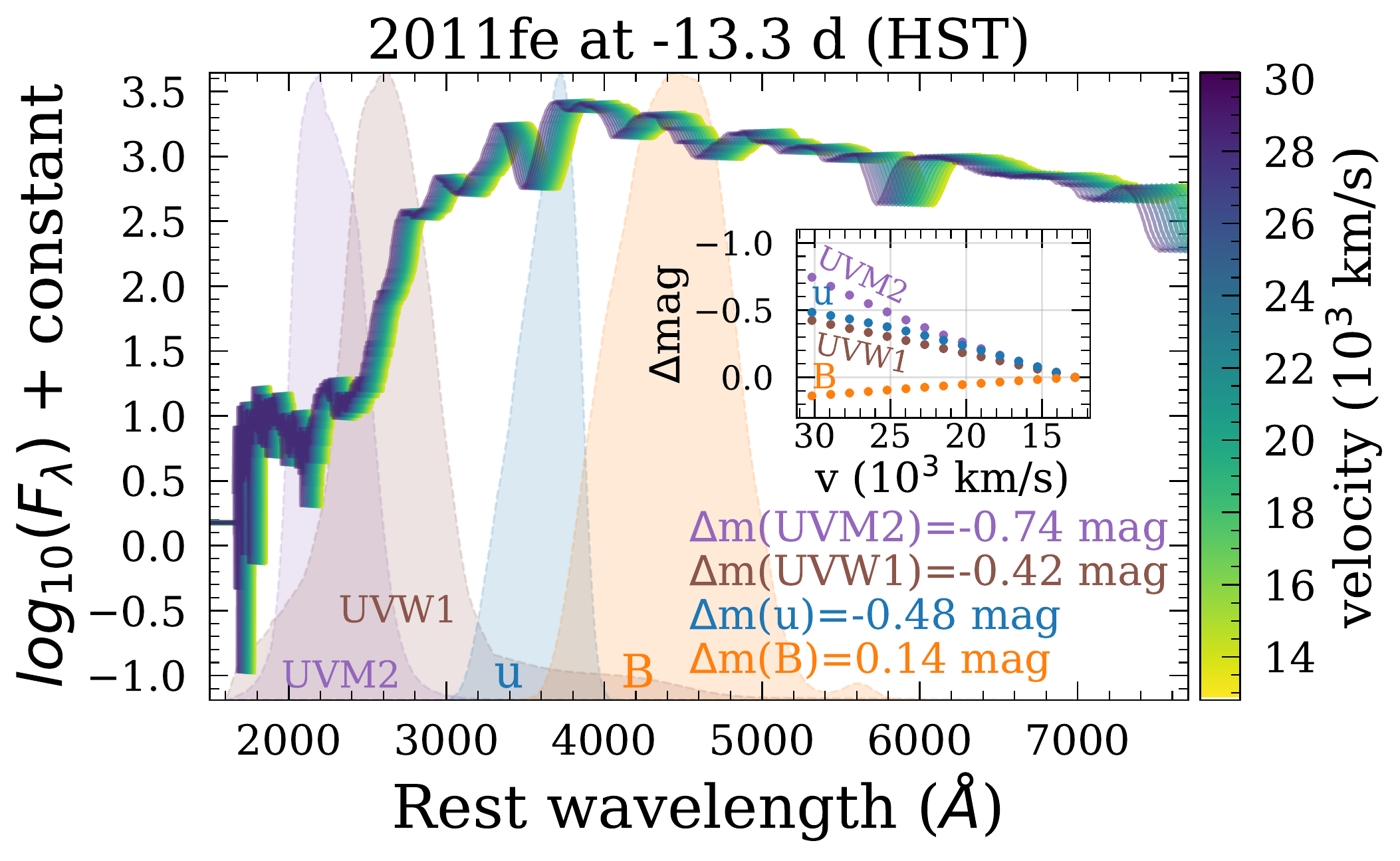}
    \end{minipage}
    \begin{minipage}[c]{.49\linewidth}
        \includegraphics[width=\linewidth]{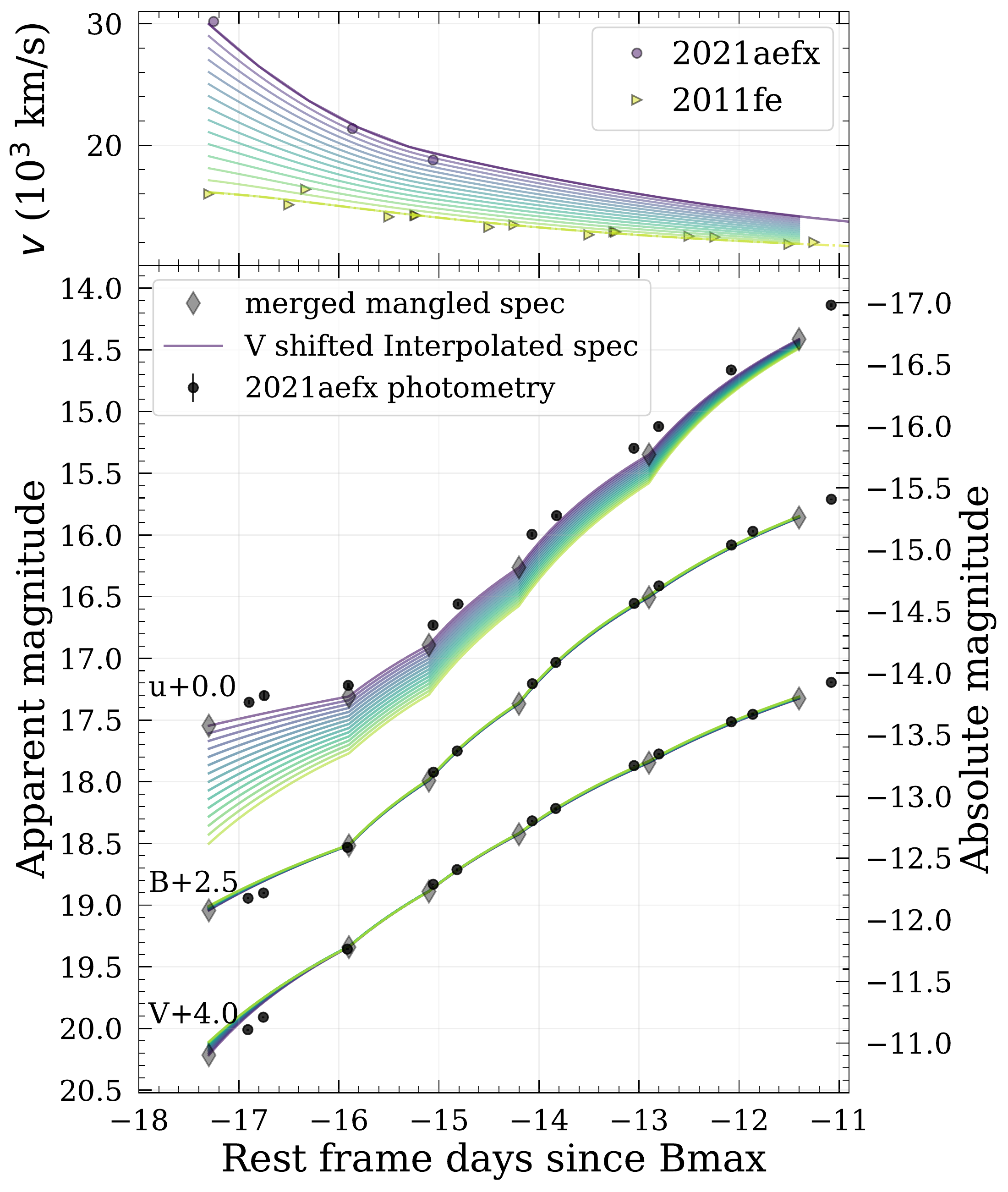} 
    \end{minipage}
    \caption{(Top left:) Illustration of how changing the velocity of the --17.3 day spectrum  can affect the $u$-band and $B$-band synthetic magnitudes. The scale on the right denotes the velocity shift, where 30,000\,\kms, is  the velocity of the observed spectra. 
    (Middle left:) Similar to the top left panel but with the spectrum of SN~2009ig at -14.2~d \citep{Foley12}. In this panel we also present the magnitude change in the swift $U$-band. 
    (Bottom left:) Similar to the top left panel but with the spectrum of SN~2011fe at -13.3~d \citep{Mazzali14} taken with the Hubble Space Telescope. Interestingly, the drop in magnitude effect is larger in the $UVM2$ band than the $u$-band, on top of this the drop in $UVW1$ is similar in magnitude to that of the $u$-band. 
    (Right:) The synthetic light curves (bottom panel) of the spectrophotometric time series with different velocity evolution.  The upper panel shows the chosen velocity evolution for this analysis. These are velocity profiles between SN~2021aefx and SN~2011fe. The $u$-band light curves show significant changes in magnitudes and shape for various velocity profiles, whereas the $B$-band remains relatively constant. At --17.3~d the $u$ band changes by  0.96~mag  between the high-velocity and low-velocity models. Conversely, the $B$-band only changes by  0.03~mag at the same epoch.}
    \label{fig:spec_with_V_shifting}
\end{figure*}

Having established that the dataset used in the analysis is spectrophotometric, we can then begin to explore if different velocity evolutions of the ejecta would produce changes in the early time behavior of the light curve. The first approach is to study the effect of velocity on the  $u$- and $B$-band photometry at a single epoch. In the top left panel of Fig. \ref{fig:spec_with_V_shifting}, the --17.3~d spectra of SN~2021aefx are presented at various velocities, shifted by increments of 1000~\kms. Note that where we discuss the velocity shift, it is relative to the minimum of the \SiII\ $\lambda$6355 feature.   At each velocity-shift the synthetic $u$- and $B$-band magnitudes are calculated.\footnote{It is important to note, that we are moving the spectra without moving the band-pass. In this case the zero point does not need recalculating. However, if the band-pass was moved the zero point would need recalculating.}
The assumption made here is that the underlying chemical distribution and state of the spectra stay the same at every velocity (i.e. the ionization state and line ratios do not change), and only a wavelength shift is applied.  The inlay in the figure demonstrates how the synthetic magnitude changes as a function of the velocity shift. Across a velocity range of 14,000~\kms,  starting at the observations (30,000~\kms) and decreasing to 16,000\,\kms, the synthetic $B$-band stays constant. However, the synthetic $u$-band drops by $\sim$1\,mag. This demonstrates that, at a given time, spectra with higher velocities are likely to have brighter $u$-band magnitudes. The drop in flux is caused by the steepness of the spectrum across the $u$ passband, specifically the emission component from the \CaII\ H\&K feature is being redshifted from the $u$- to the $B$-band as the velocity of the spectrum decreases. In the middle left panel of  Fig. \ref{fig:spec_with_V_shifting} we also show the same velocity test carried out with the $-14.2$\,d spectrum of SN~2009ig. The difference in in the $u$-band magnitude between the low velocity and high velocity spectrum is 0.87~mag.

The next stage of the analysis is to determine if various velocity evolution pathways of the spectral time series significantly affect the rising flux of the SN.
To do this, a range of velocity profiles was created between the evolution of SN~2021aefx and SN~2011fe; see the right panel of Fig. \ref{fig:spec_with_V_shifting}. 
There is no significant difference between the low-velocity and high-velocity profiles in the $B$ band, however there is a clear difference in the early $u$-band behavior. The shape of the rising light curve in $u$ changes dramatically within the first few days between the high-velocity and low-velocity time series.  The high-velocity time series produces  an apparent  early-excess flux within the $u$-band. 
If the different synthetic $u$-bands are fit with a power law,
the high velocity time series requires a two-component power law fit, whereas the low-velocity time series can be fit by a power law of f $\propto$ t$^{7.7}$.

In the $u$-band the low-velocity time series is 0.96~mag fainter than the high-velocity time series (i.e. the observations) at --17.3\,d. The  low-velocity time series rises smoothly and is similar to that of a `normal' SN\,Ia light curve. However, the high-velocity time series has an early-excess flux in the $u$ band. These results  suggest that in this case, this excess flux is caused by line profiles shifting to shorter wavelengths. In particular,  at early times the \CaII\ H\&K emission is located within the $u$ band for SNe that have high velocities, but not for SNe~Ia that have lower velocities. An example of how different velocity gradients affects the underlying opacity can be seen in Fig. 1 \& 2 of \citet{Hoeflich93}.

Many SNe~Ia that have early-excess emission in the $u$-band also have excess-emission at mid-UV wavelengths. SN~2021aefx does not have early time UV observations available, therefore we cannot test the effects directly for this object. However,  as a test case we use the -13.3~d spectrum of SN~2011fe to determine if the rapid drop in spectral flux can also produce early-excess flux in the UV (see the bottom left panel of Fig. \ref{fig:spec_with_V_shifting}). For SN~2011fe the Swift $UVM2$ and $UVW1$ magnitudes drops by 0.74~mag and 0.41~mag respectively over a velocity change from 30,000\,\kms\ to 13,000\,\kms, but stays constant in the $B$-band. We note that the HST spectra does not go far enough into the blue to test that $UVw2$ filter. While the spectrum of SN~2011fe is taken at a different time compared to the phase where the blue early-excess is seen, it acts as a proof of concept. Generally, if there is a large slope in the flux across a passband then a rapid change in velocity of the profiles can cause additional structure within the rising light curves.  However, we note that there is another effect of velocities at these mid-UV wavelength ranges, since the opacity in this region is almost entirely dominated by line blanketing due to iron group elements \citep{DerKacy20}. Increasing velocity tends to depress the flux in the UV since the increased velocity leads to an increased effective opacity in this region \citep{Wang12}, which our assumptions neglects. This effect is clearly seen in SNe 2017erp and 2021fxy (J.M.~DerKacy et al., in preparation).

Finally, we tested how the $u$-band flux changes with spectra of SNe~Ia which do not have  broad spectral features (see the bottom left panel of  Fig. \ref{fig:spec_with_V_shifting}). By using the spectrum of SN~2011fe at --13.3~d the minimum of the features were artificially increased in velocity to 30,000~\kms. The change in $u$-band magnitude between the high-velocity spectrum and low-velocity spectrum is 0.5\,mag demonstrates that velocity has an important effect. However, the effect of the velocity shift is smaller in SN~2011fe than in SN~2021aefx because the features in SN~2011fe are narrower. For SN~2021aefx with its broader features, the short wavelength edge of the absorption of \CaII\ H\&K will \textit{eat into} the emission region of the adjacent shorter wavelength features, producing excess line blanketing and further reducing the flux in the region. Therefore, we conclude the effect seen in SN~2021aefx is produced by both the high-velocity and the broadness of the spectral features.

After the submission of this work \citet{Hosseinzadeh22} presented early time observations of SN~2021aefx. They showed that early-time open-filter light curve of SN 2021aefx had structure which may be interpreted as an early-excess flux. However, the early-time shoulder in the DLT40 open filter is different from the early-excess blue flux see in the $u$ and $UV$ filters, but is similar in shape our $V$-band light curves. In this work we concentrate on the nature of the $u$ and $UV$ early flux as this shows the largest early-excess flux. Our work here demonstrates that for this particular SN, the Doppler effect accounts for most of the excess in the $UV$ and $u$-bands leaving little room for additional effects. Therefore, any model fitting of early SNe Ia must consider the effect of Doppler shift on the $u$ and $UV$ light curves.

\section{Discussion} \label{sect:Discussion}

We provide a cautionary tale that the early time diversity seen in $u$-band light curves of SNe~Ia can be caused, in part, by the rapid change in the velocity of the Ca II H+K feature. This prediction is only true for SNe~Ia which have early-time high-velocity and broad spectral profiles (i.e. objects which have expansion velocities higher than ~25,000\kms\ in the earliest phases).

The effect is produced by the \CaII\ H\&K emission rapidly passing from the $u$ band and into the $B$ band causing a change in flux and change in slope of the $u$-band light curve.  The rapid change in velocity can produce various $u$-band light curves which monotonically increase, but it cannot explain all of the early time diversity in SNe~Ia light curves, such as the more complex bump-like shapes seen in objects like SN~2017cbv \citep{Hosseinzadeh17}. In fact, given the fact that SN~2017cbv has ejecta velocities lower than $\eqsim$25,000~\kms\ we would not expect  the effect discussed here to be the dominant cause. 
Furthermore, in normal velocity SNe~Ia such as SN~2011fe this effect is also not expected to be large. However, for a light curve of an object such as SN~2012cg \citep{Marion16} the effect discussed here may be able to  explain the early blue excess flux. Only spectra taken in this earliest phase would reveal if the lines are of high enough velocity or broad enough for the effect to be dominant. 

The work here should also provide a word of warning that if one is to 
interpret the $u$-band flux between observations at different telescopes of the same SN, a precise knowledge of the transmission function is required otherwise variation between $u$-band flux may be due to whether the blue edge of the \CaII\ H\&K feature is in the bandpass or not.

To fully understand the effects that a rapid change in velocity has on the light curve requires high-cadence spectral data which cover the whole of the rest-frame $u$ band, and are obtained within hours after the explosion. Future follow-up observations of transients should ensure that the whole wavelength range is observed. 

Finally, although this work may not explain all of the diversity within early SNe~Ia light curves, our results should be carefully considered when interpreting physical meaning from SNe~Ia which have early-time high ejecta velocities and early-excess UV and $u$-band light curves.

\section{Acknowledgments}
We thank the referee for their helpful and constructive comments. 
C.A. and B.J.S. are supported by NSF grants AST-1907570, AST-1908952, AST-1920392,
and AST-1911074.
M.D.S. of funded in part by an Experiment grant (\# 28021) from the Villum FONDEN, and by a project 1 grant (\#8021-00170B) from the Independent Research Fund Denmark (IRFD)
PH acknowledge support
by National Science Foundation (NSF) grant AST-
1715133.
E.B. and J.D. are supported in part by NASA grant
80NSSC20K0538.
This work has been generously supported by the National Science Foundation under grants AST-1008343, AST-1613426, AST-1613455, and AST1613472.
This paper includes data gathered with the 6.5 meter Magellan Telescopes located at Las Campanas Observatory, Chile.
We would like to thank the technical staff for constant support for observations on the Swope telescope.
 The early time spectrum which was critical for this analysis came from SALT through Rutgers University time via program 2021-1-MLT-007 (PI: Jha). 
This paper includes data gathered with the 6.5 meter Magellan Telescopes located at Las Campanas Observatory, Chile
 L.G. acknowledges financial support from the Spanish Ministerio de Ciencia e Innovaci\'on (MCIN), the Agencia Estatal de Investigaci\'on (AEI) 10.13039/501100011033, and the European Social Fund (ESF) "Investing in your future" under the 2019 Ram\'on y Cajal program RYC2019-027683-I and the PID2020-115253GA-I00 HOSTFLOWS project, from Centro Superior de Investigaciones Cient\'ificas (CSIC) under the PIE project 20215AT016, and the program Unidad de Excelencia Mar\'ia de Maeztu CEX2020-001058-M.

\bibliographystyle{aasjournal}
\bibliography{bibtex} 




\end{document}